\documentclass[11pt,preprintnumbers,aps,amssymb,nofootinbib,amsmath,superscriptaddress]
{revtex4}
\usepackage{epsfig,epsf}
\usepackage{bm} 
\usepackage{color}
\newcommand{\beq}{\begin{equation}}
\newcommand{\beql}[1]{\begin{equation}\label{#1}}
\newcommand{\eeq}{\end{equation}}
\def\bal#1\gal{\begin{align}#1\end{align}}
\newcommand{\ball}[1]{\bal\label{#1}}
\newcommand{\eq}[1]{(\ref{#1})}
\newcommand{\fig}[1]{Fig.~\ref{#1}}
\renewcommand{\sec}[1]{Sec.~\ref{#1}}
\newcounter{topiccounter}
\renewcommand{\b}[1]{{\bm #1}} 
\newcommand{\unit}[1]{\hat {{\bm #1}}} 

\newcommand{\im}{\,\mathrm{Im}\,}

\newcommand{\aver}[1]{\left\langle #1 \right\rangle}

\begin{document}


\title{Electromagnetic field and the chiral magnetic effect in the quark-gluon plasma }

\author{Kirill Tuchin}

\affiliation{
Department of Physics and Astronomy, Iowa State University, Ames, IA 50011}

\date{\today}


\begin{abstract}

Time evolution of an electromagnetic field created in heavy-ion collisions  strongly depends on the  electromagnetic response of the quark-gluon plasma, which can be described by the Ohmic and chiral conductivities. The later is intimately related to the Chiral Magnetic Effect. I argue that a solution to the classical Maxwell equations at finite chiral conductivity is unstable due to the soft modes $k<\sigma_\chi$ that  grow exponentially with time. In the kinematical region relevant for the relativistic heavy-ion collisions,  I derive analytical expressions for the magnetic field of a point charge. I show that finite chiral conductivity causes oscillations of magnetic field at early times.

\end{abstract}

\maketitle

\section{Introduction}\label{sec:a}

Collision of relativistic heavy-ions produces hot nuclear matter that can be described using the relativistic hydrodynamics \cite{Landau:1953gs,Belenkij:1956cd}. I will refer to this matter as the Quark-Gluon Plasma (QGP) leaving aside the issues of its equilibration and thermalization. Valence electric charges of the colliding ions are not a part of the plasma, as they continue on the incident trajectory along the beam directions with very little deflection \cite{Itakura:2003jp}. However, they create strong electromagnetic field (EMF) that influences the plasma behavior \cite{Kharzeev:2007jp,Tuchin:2013ie,Voronyuk:2011jd,Skokov:2009qp,Deng:2012pc,Bzdak:2011yy}. Electrically conducting plasma responds by generating induced EMF. The resulting EMF is a solution to a complicated magneto-hydrodynamic problem. As a first approximation, one can rely on  slow time-dependence  of the relevant kinetic coefficients on time to decouple the Maxwell equations from the time evolution of the QGP. Analytical solution to these equations shows that the EMF decreases with time much slower than in vacuum and is approximately collision energy independent; rather it depends only on the impact parameter and the electrical conductivity of the QGP \cite{Tuchin:2010vs,Tuchin:2013apa,Tuchin:2013ie,Gursoy:2014aka}. Numerical simulations that take into account the QGP expansion \cite{Zakharov:2014dia} qualitatively agree with this conclusion.\footnote{A different strength of EMF in \cite{Zakharov:2014dia} and \cite{Tuchin:2013apa} is due to different initial time at which the plasma evolution starts.} 

It has been recently realized that kinetic properties of the QGP reflect the nontrivial topological structure of the QCD. In particular, the QGP responds to the chirality imbalance by generating metastable parity-odd domains. In the presence of external magnetic field such a metastable domain induces a parallel to it electric field, which is known as the Chiral Magnetic Effect (CME) \cite{Kharzeev:2004ey,Kharzeev:2007jp,Kharzeev:2007tn,Fukushima:2008xe,Kharzeev:2009fn}. Electric current generated by the CME is proportional to the external magnetic field, with the chiral conductivity $\sigma_\chi$ being the proportionality coefficient. In this paper, I study the electromagnetic field generated by valence charges at finite chiral conductivity and determine the role of the Chiral Magnetic Effect (CME)  in the electromagnetic field dynamics in the QGP. 

I found a two-fold effect of the CME on the electromagnetic field evolution. Firstly, the field becomes unstable because soft modes with $k<\sigma_\chi$ grow exponentially with time. For the QGP this effects is of little importance since the largest wavelength $1/k$ that is allowed in QGP is much smaller than $1/\sigma_\chi$. However, in non-Abelian plasmas with large spatial extent this is an  important phenomenon that may lead to a breakdown of electromagnetic field into a set of knots with non-trivial topology.\footnote{A different type of  ``chiral plasma instabilities" has been recently discussed in \cite{Joyce:1997uy,Boyarsky:2011uy,Tashiro:2012mf,Grabowska:2014efa,Akamatsu:2013pjd,Akamatsu:2014yza}.} Secondly, due to finite chiral conductivity, magnetic field, produced by valence electric charges, oscillates at early times after a heavy-ion collision. These oscillations  may result in partial cancelation of the magnetic field effects, when averaged over time.

The paper is structured as follows: In \sec{sec:b} I describe the Maxwell-Chern-Simons (MCS) theory, which is an elegant way to incorporate the topological effects in QED.  In the MCS the  chiral conductivity arises from the time-dependent $\theta$-angle. Following  \cite{McLerran:2013hla} I consider a simplest model with constant   $\sigma_\chi$.  In \sec{sec:c} I solve MCS equations away from charges and show that the dispersion relation of electromagnetic wave contains an unstable mode at $k<\sigma_\chi$. In \sec{sec:d} I derive expressions for the electromagnetic field of a relativistic point charge and discuss its properties.  Explicit analytical expressions for the magnetic field of a point charge is derived in \sec{sec:g} in the diffusion approximation, which is appropriate for the relativistic heavy-ion collisions. The main result,  shown in \fig{fig2},  indicates that at finite chiral conductivity, magnetic field components oscillate at early times. I discuss these results and conclude in \sec{sec:i}.

\section{ Maxwell-Chern-Simons equations}\label{sec:b}

The Lagrangian of electrodynamics coupled to the topological charge carried by the gluon field, the so-called Maxwell-Chern-Simons theory, reads \cite{Wilczek:1987mv,Carroll:1989vb, Sikivie:1984yz,Kharzeev:2009fn} 
\ball{b10}
L= -\frac{1}{4}F^{\mu\nu}F_{\mu\nu}-A_\mu j^{\mu}-\frac{c}{4}\theta\tilde F^{\mu\nu}F_{\mu\nu}\,,
\gal
where $c=N_c\sum_fq_f^2e^2/2\pi^2$.  An external pseudo-scalar field $\theta$ depends on the medium properties and originates in the QCD Lagrangian. The corresponding field equations are given by  \footnote{   The correct signs in front of the anomalous terms where derived in \cite{Ozonder:2010zy}. } 
\bal
&\b \nabla\cdot \b B=0\,, \label{b11}\\
& \b \nabla\cdot \b E= \rho- c\, \b\nabla\theta\cdot \b B\,,  \label{b12}\\
& \b \nabla \times \b E= -\partial_t \b B\,,\label{b13}\\
& \b \nabla \times \b B= \partial_t \b E+ \b j + c(\partial_t\theta\, \b B+ \b\nabla \theta\times \b E)\,.\label{b14}
\gal
Time-derivative $\dot \theta= \mu_5$ can be identified with the axial chemical potential $\mu_5$ \cite{Fukushima:2008xe,Kharzeev:2009fn}. The part of the anomalous current density proportional to the magnetic field can be written down as $\b j= \sigma_\chi \b B$, where 
\ball{b25}
\sigma_\chi = \mu_5 \frac{e^2}{2\pi^2}N_c\sum_fq_f^2\,
\gal
is the chiral conductivity induced by the QED anomaly \cite{Kharzeev:2009pj}. The $\theta$-angle is believed to be finite inside metastable regions of size $\sim 1/g^2T$. On average it must vanish $\aver{\theta}=0$ to preserve the global $\mathcal{CP}$-invariance of the QCD. Its space and time dynamics is complicated: shortly after a heavy-ion collision it is determined by the colored fields of glasma  \cite{Kharzeev:2001ev,Lappi:2006fp,Hirono:2014oda}, while at later time by the sphaleron transition dynamics \cite{Joyce:1997uy,Boyarsky:2011uy,Tashiro:2012mf,Grabowska:2014efa}.

Since the detailed structure of inhomogeneous field $\theta$ is unknown, one has to resort to phenomenological models in order to study its effect on the electromagnetic field dynamics (see e.g.\  \cite{Hirono:2014oda}). The simplest model that captures the essential dynamics of the CME effect, and that we adopt in the present study, is to neglect the space variation of $\theta$ and approximate $\sigma_\chi$ by a constant. In other words we set $\b\nabla\theta=0$ and $\sigma_\chi=\text{const}$. This model was used in \cite{Chernodub:2010ye} to discuss non-trivial static topological solutions of \eq{b11}--\eq{b14} (see below) and in \cite{McLerran:2013hla} to numerically investigate time-evolution of magnetic field. The main advantage of this model is that it can be analytically solved  and thus provides important insights into the dynamics of the electromagnetic fields in the presence of the chiral anomaly. Moreover, it is argued in \cite{Zhitnitsky:2014ria,Zhitnitsky:2014dra} that $\theta$ may actually be  a slow function of $x$ that permits expansion $\theta\approx \theta_0+ \mu_5 t + c^{-1}\b P\cdot \b r$ with constant $\mu_5$ and $\b P$. 

Consider now the system of equations \eq{b11}--\eq{b14} in the absence of electric charges, with the assumptions discussed in the previous paragraph. It  has non-trivial stationary solutions with finite magnetic field and vanishing electric field that satisfies the following equations \cite{Chandra,RB,CDGT}:
\bal
&\b \nabla\cdot \b B=0\,, \label{z11}\\
&\b \nabla \times \b B=  \sigma_\chi \b B\,. \label{z12}
\gal
It is argued in \cite{Chernodub:2010ye} that since the anomalous current $\b j =\sigma_\chi \b B$ exists only in the deconfined phase occupying  a domain of finite volume $D$, there is no outward current on its boundary. This implies the boundary condition 
\ball{z14}
\unit r \cdot \b B\big|_{\partial D} =0\,.
\gal
Solution to \eq{z11}-\eq{z14} is a system of magnetized knots of different sizes. In a simplest case of spherical boundary the possible values of its  radius are
\ball{z16}
R_n= \frac{\kappa_n}{\sigma_\chi}\,,\quad n=0,1,2,\ldots\ldots\,,
\gal
where $n$ enumerates zeros of spherical Bessel functions $\kappa_n$. The smallest  of $\kappa$'s is $\kappa_0\approx 4.5$, which for a realistic $\sigma_\chi$ yields $R_0\approx 200$~fm. $R_0$ is much larger than a characteristic transverse size of the QGP $R_A\sim 6-10$~fm and thus has no effect on the QGP phenomenology. It is possible that magnetic knots are artifacts of our model for the $\theta$-angle. It is far from clear whether any static topological solutions survive in a more realistic model.

\section{Instability of Electromagnetic waves in infinite plasma}\label{sec:c}

Consider electromagnetic waves  propagating in plasma far from any sources. In a conducting medium Maxwell equations for the electromagnetic field  read
\bal
&\b \nabla\cdot \b B=0\,, \label{c11}\\
& \b \nabla\cdot \b D= 0\,,  \label{c12}\\
& \b \nabla \times \b E= -\partial_t \b B\,,\label{c13}\\
& \b \nabla \times \b H= \partial_t \b D + \sigma_\chi \b B\,.\label{c14}
\gal
$\b D$ is electric displacement vector. We will assume that $\mu =1$. Fourier transformation 
\ball{c19}
\b E(\b r,t)= \int \frac{d^4k}{(2\pi)^4}e^{-ik\cdot x}\b E_{\omega,\b k}\,,\quad \b B(\b r,t)= \int \frac{d^4k}{(2\pi)^4}e^{-ik\cdot x}\b B_{\omega,\b k}
\gal
where $x=(t,\b r)$, $k= (\omega, \b k)$ yields  Maxwell  equations in momentum space
\bal
& \b k\cdot \b B_{\omega,\b k}= 0\,, \label{c21}\\
&\epsilon \b k\cdot \b E_{\omega,\b k}= 0\,, \label{c22}\\
& \b k\times \b E_{\omega,\b k}= \omega \b B_{\omega,\b k}\,, \label{c23}\\
& \b k\times \b B_{\omega,\b k}= -\omega \epsilon \b E_{\omega,\b k}-i\sigma_\chi \b B_{\omega,\b k}\,, \label{c24}
\gal
where $\b D_{\omega,\b k}= \epsilon \b E_{\omega,\b k}$. In electrically conducting medium  with the Ohmic conductivity $\sigma$ the permittivity is  $\epsilon= 1+i\sigma/\omega$,
Taking  vector product of \eq{c24} with $\b k$ and using \eq{c21} and \eq{c23} we get
\ball{c27}
\b B_{\omega,\b k}[\omega(\omega+i\sigma)-\b k^2]= -i\sigma_\chi \b k\times \b B_{\omega,\b k}\,.
\gal
Taking another vector product with $\b k$ gives
\ball{c29}
(\b k\times \b B_{\omega,\b k})[\omega(\omega+i\sigma)-\b k^2]= i\sigma_\chi \b k^2 \b B_{\omega,\b k}\,.
\gal
Equations \eq{c27} and \eq{c29} have a non-trivial solution only if the following dispersion relation is satisfied
\ball{c31}
[\omega(\omega+i\sigma)-\b k^2]^2=\sigma_\chi^2\b k^2\,.
\gal
It has four solutions
\ball{c33}
\omega_{\lambda_1,\lambda_2}= -\frac{i\sigma}{2}+\lambda_1\sqrt{ k^2+\lambda_2\sigma_\chi k- \sigma^2/4}\,,
\gal
where $\lambda_1,\, \lambda_2= \pm 1$  and $k=\sqrt{\b k^2}\ge 0$. These solutions determine the time dependence of electromagnetic wave as $\sim e^{-i\omega_{\lambda_1,\lambda_2}t}$.

Let $\kappa^2= k^2+\lambda_2\sigma_\chi k- \sigma^2/4$.
When $\kappa^2>0$ the electromagnetic wave oscillates with frequency $\kappa$ and is damped over the distance $1/\sigma$. This corresponds to momenta 
\ball{c35}
k>k_0\equiv \frac{1}{2}\sqrt{\sigma_\chi^2+\sigma^2}-\frac{\lambda_2\sigma_\chi}{2}\,.
\gal
For $k<k_0$, $\kappa^2<0$, and all $\omega_{\lambda_1,\lambda_2}$'s  become imaginary implying that electromagnetic wave is a monotonic function of time. At $\kappa^2=-\sigma^2/4$, which occurs at $k=\sigma_\chi$, $\lambda_2=-1$, and $\lambda_1=+1$, $\omega_{+-}$ vanishes indicating a stationary mode. Finally, when $\kappa^2<-\sigma^2/4$, i.e.
 $k<\sigma_\chi$, $\lambda_2=-1$, $\lambda_1=+1$ there is an unstable mode with $\im \omega_{+-}>0$ which corresponds to  exponentially increasing magnetic field.  $\im \omega_{+-}$ vanishes at $k=0$ and $k=\sigma_\chi$ and has a maximum value of $\left(\sqrt{\sigma^2+\sigma_\chi^2}-\sigma\right)/2$ at $k=\sigma_\chi/2$.

Electromagnetic wave which at some initial time contains modes extending to the region $k<\sigma_\chi$ is unstable. This is a usual situation in an infinite plasma. However, in a plasma of spatial size $R$ there are only modes $k\gtrsim 1/R$. Therefore, the instability affects the field evolution only if $R\gtrsim1/\sigma_\chi$. In the QGP this condition is not satisfied, except, perhaps, in a very rear fluctuations of the $\theta$-angle, and hence can be ignored.

\section{Electromagnetic field of a point charge}\label{sec:d}

In electrically  conducting medium Maxwell equations for the electromagnetic field  of a point charge moving along a straight line $z=vt$ read
\bal
&\b \nabla\cdot \b B=0\,, \label{d11}\\
& \b \nabla\cdot \b D= e\delta (z-vt)\delta (\b b)\,,  \label{d12}\\
& \b \nabla \times \b E= -\partial_t \b B\,,\label{d13}\\
& \b \nabla \times \b H= \partial_t \b D + \sigma_\chi \b B+ ev\unit z \delta (z-vt)\delta (\b b)\,.\label{d14}
\gal
These equations in momentum space are
\bal
& \b k\cdot \b B_{\omega,\b k}= 0\,, \label{d21}\\
&\epsilon \b k\cdot \b E_{\omega,\b k}= -2\pi i e\delta (\omega - k_z v)\,, \label{d22}\\
& \b k\times \b E_{\omega,\b k}= \omega \b B_{\omega,\b k}\,, \label{d23}\\
& \b k\times \b B_{\omega,\b k}= -\omega \epsilon \b E_{\omega,\b k}-i\sigma_\chi \b B_{\omega,\b k}-2\pi i ev\unit z  \delta(\omega-k_zv)\,. \label{d24}
\gal
We repeat the algebraic manipulations  of the previous section. Firstly, taking the vector product of \eq{d24} with $\b k$ and using \eq{d21} and \eq{d23} we arrive at 
\ball{d27}
\b B_{\omega,\b k}[\omega(\omega+i\sigma)-\b k^2]= -i\sigma_\chi \b k\times \b B_{\omega,\b k}-2\pi i ev\b k \times \unit z  \delta(\omega-k_zv)\,.
\gal
Secondly, we take another vector product with $\b k$ to obtain
\ball{d29}
(\b k\times \b B_{\omega,\b k})[\omega(\omega+i\sigma)-\b k^2]= i\sigma_\chi \b k^2 \b B_{\omega,\b k}-2\pi i e v \b k\times (\b k\times \unit z)\delta(\omega-k_zv)\,.
\gal
We are interested in a particular solution to equations \eq{d27},\eq{d29}, namely the one  that is generated by the electric charge $e$.  Solving \eq{d27} and \eq{d29} yields 
\ball{d31}
\b B_{\omega,\b k}= \frac{(\b k\times \unit z) [\omega(\omega+i\sigma)-\b k^2]-i\sigma_\chi \b k\times (\b k\times \unit z)}{[\omega(\omega+i\sigma)-\b k^2]^2-\sigma_\chi^2\b k^2}(-2\pi i) ev\delta(\omega - k_z v)\,.
\gal
Electric field follows from the Faraday law \eq{d23} upon taking its vector product with $\b k$:
\ball{d33}
\b k (\b k\cdot \b E_{\omega,\b k})-\b k^2 \b E_{\omega,\b k}= \omega (\b k\times \b B_{\omega,\b k})\,.
\gal
Substituting \eq{d22} and \eq{d24} we  find 
\ball{d35}
\b E_{\omega,\b k}= \frac{2\pi  ie\delta(\omega - k_z v)[\b k/\epsilon-v\omega \unit z]-i\omega \sigma_\chi \b B_{\omega,\b k}}{\omega(\omega+i\sigma)-\b k^2}\,,
\gal
with $\b B_{\omega,\b k}$ given by \eq{d31}.

It will be suitable to write the cross products in \eq{d31} in cylindrical coordinates. Let $\psi$ be the angle between the vector $\b k_\bot$ and the $x$-axis, the corresponding unit vector is $\unit \psi=-\unit x \sin\psi+\unit y \cos\psi$. Then 
\bal
&\b k\times \unit z= -k_\bot \unit \psi\,, \label{d37}\\
& \b k\times(\b k\times \unit z)= k_z \b k_\bot -k_\bot^2\unit z\,. \label{d38}
\gal 
Using identities \eq{d37},\eq{d38} in  \eq{d22}, substituting the result into  \eq{c19} and taking integral over $k_z$ we find 
\ball{d40}
\b B= ie\int_{-\infty}^{+\infty}\frac{d\omega}{2\pi}\int \frac{d^2k_\bot}{(2\pi)^2}
\frac{k_\bot \unit \psi [\omega(\omega+i\sigma)- k_\bot^2-\frac{\omega^2}{v^2}]+i\sigma_\chi (\b k_\bot \frac{\omega}{v}-k_\bot^2\unit z)}{[\omega(\omega+i\sigma)- k_\bot^2-\frac{\omega^2}{v^2}]^2-\sigma_\chi^2(k_\bot^2+\frac{\omega^2}{v^2})}e^{-i\omega x_-+i \b k_\bot \cdot \b b}\,.
\gal
where $x_-= t-z/v$.

\begin{figure}[ht]
      \includegraphics[height=5cm]{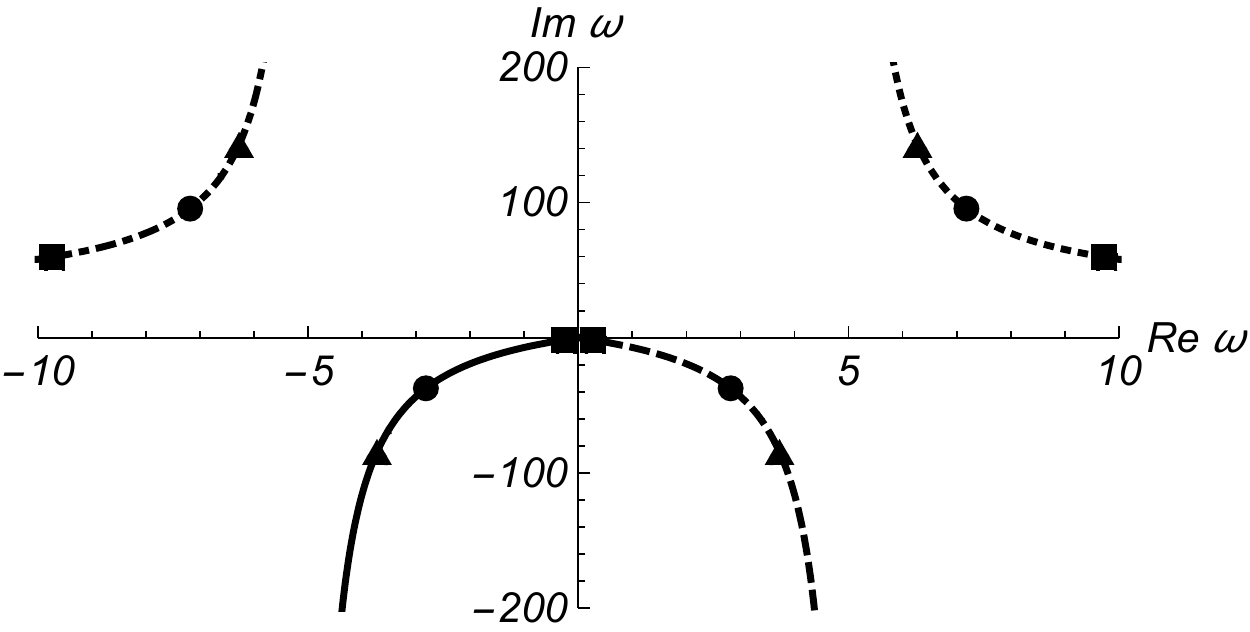} 
  \caption{Four solutions of \eq{d59} at $\sigma=5.8$~MeV, $\sigma_\chi=1$~MeV,  $\gamma=100$. Horizontal and vertical axes are in units of GeV. Each line is a unique function of $k_\bot$. Squares, circles and triangles indicate the positions of the poles at $k_\bot=0.1,0.6,1.1$~GeV respectively. }
\label{fig1}
\end{figure}

Time dependence of magnetic field is determined by the poles of \eq{d31} in the plane of complex $\omega$. These poles are solutions of the following quartic equation 
\ball{d57}
\left[\omega(\omega+i\sigma)- k_\bot^2-\frac{\omega^2}{v^2}\right]^2-\sigma_\chi^2\left(k_\bot^2+\frac{\omega^2}{v^2}\right)=0\,.
\gal
Eq.~\eq{d57} can be obtained from the dispersion relation \eq{c31} of a free wave by restricting it to  the particle equation of motion $k_z= \omega/v$.  Introducing $\gamma = (1-v^2)^{-1/2}$ allows us to cast \eq{d57} in a more convenient form
\ball{d59}
\left( -\frac{\omega^2}{v^2\gamma^2}+i\omega \sigma - k_\bot^2\right)^2-\sigma_\chi^2\left( \frac{\omega^2}{v^2}+k_\bot^2\right)=0\,.
\gal
Four solutions to this equation can be found using the standard algebraic methods. However, they are quite bulky, so I am not reproducing them here.  Instead, I find it more illuminating to plot them  at fixed $\sigma$, $\sigma_\chi$ and $\gamma$  for different values of $k_\bot$ as shown in \fig{fig1}.

Position of the four poles at $k_\bot\to 0$ can be found by expanding  \eq{d59}, which gives three distinct solutions  $\omega=0$ and $\omega=v^2\gamma^2(i\sigma\pm \sigma_\chi)$. The former corresponds to the minimum value of the lower branches, while the later  to the minimum values of the upper branches.
Thus, the upper branches are separated from the real axis by a gap $v^2\gamma^2\sigma$. The absolute value of the real part of the upper branches decreases monotonically with $k_\bot$. At $k_\bot\to \infty$
\ball{d60}
\omega\approx\pm iv\gamma k_\bot \pm\frac{1}{2}v\gamma\sigma_\chi\sqrt{\gamma^2-1}\,,
\gal
Thus, the real value of $\omega$ of upper branches approaches a constant at large $k_\bot$, which indicates that a gap of size $\sim \gamma^2 \sigma_\chi$  exists also between the upper branches and the imaginary axis. In the ultra-relativistic limit $v\to 1$, or $\gamma\to \infty$, the upper branches move to infinity. Since the poles in the upper-half plane determine the electromagnetic field at  $x_-<0$, it gets exponentially suppressed at $\gamma\gg 1$.

Behavior of the electromagnetic field at $x_->0$ is determined by the two poles in the lower half-plane. Unlike the poles in the upper half-plane they stay finite in the ultra-relativistic limit. One of the lower branches exhibits a peculiar behavior by crossing the real axis and acquiring a positive $\im \omega$ when $k_\bot<\sigma_\chi$. This is a way  in which the field instability discussed in the previous section  manifests itself in this case. (This feature is not readily seen in \fig{fig1}  due the small value  of $\sigma_\chi$).  Existence of a pole in the upper-half plane implies that the field of a point charge moving along $x_-=0$ receives acausal contribution, viz.\ a term that is finite at $x_-<0$  when $\gamma\to \infty$.  Fortunately, transverse momenta as small as $k_\bot\sim \sigma_\chi$ are not relevant in relativistic heavy-ion phenomenology allowing me to neglect the acausal contribution. This however does not resolve a theoretical problem that the acausal term presents.\footnote{A solution to this problem might be related to existence magnetic  knots discussed in \sec{sec:b} that also appear at $k\sim \sigma_\chi$.  }

\section{Diffusion approximation}\label{sec:g}

At a given light-cone time $x_->0$ the $\omega$-integral in \eq{d40} vanishes at $\omega\gg 1/x_-$ due to the rapid oscillation of the integrand. Therefore, at later times the terms in \eq{d59} that are   quadratic in $\omega$ are suppressed. This correspond to the following ``diffusion" approximation:
\ball{g11}
 \omega\ll \sigma v^2\gamma^2\,,\quad \omega\ll v\gamma k_\bot\,,
\gal    
which is tantamount to
\ball{g13}
 x_-\gg \frac{1}{\sigma v^2\gamma^2}\,,\quad x_-\gg \frac{b}{v\gamma}\,,
\gal
where we estimated $k_\bot\sim 1/b$. Electrical conductivity of the quark-gluon plasma at the critical temperature is $\sigma = 5.8$~MeV \cite{Ding:2010ga,Aarts:2007wj,Amato:2013oja,Cassing:2013iz}. For a heavy-ion collision at $\gamma=100$ we estimate $1/\sigma v^2 \gamma^2\sim 3\cdot 10^{-3}$~fm. For $b\sim 10$~fm, $b/\gamma \sim 0.1$~fm. Taking into account that it takes about $1/Q_s\sim 0.2$~fm to release the color charges from the nuclei wave functions, it follows that approximation \eq{g11} applies to the entire lifetime of the QGP. The precise initial conditions do not play an important role in the electromagnetic field evolution. 

Since the valence quarks are ultra-relativistic, i.e.\ $\gamma\gg 1$, we will approximate their velocity as $v\approx 1-1/2\gamma^2$. Then, the dispersion relation \eq{d59}  in the diffusion approximation takes form
\ball{g15}
\left( i\omega \sigma - k_\bot^2\right)^2-\sigma_\chi^2\left( \omega^2+k_\bot^2\right)=0\,.
\gal
The two solutions of \eq{g15}, describing the two lower poles in \fig{fig1},  are 
\ball{g17}
\omega_{1,2}= \frac{-i\sigma k_\bot^2\pm k_\bot \sigma_\chi \sqrt{k_\bot^2-\sigma^2-\sigma_\chi^2}}{\sigma^2+\sigma_\chi^2}\,.
\gal 
 These are the only poles of the Fourier component of magnetic field $\b B_{\omega,\b k}$ in the complex $\omega$-plane because the upper poles in \fig{fig1} disappear in the limit $v\to 1$. If $k_\bot> \sqrt{\sigma^2+\sigma_\chi^2}$, then both complex-conjugated poles lie in the lower half-plane. If $\sigma_\chi<k_\bot< \sqrt{\sigma^2+\sigma_\chi^2}$, then there are two  poles on the imaginary axis in the lower half-plane. Finally, if $k_\bot<\sigma_\chi$, then both poles lie on the imaginary axis, but $\omega_1$ is in the upper-half plane, while $\omega_2$ is still in the lower one. 

In the diffusion approximation \eq{d40} reads
\bal
\b B&= -ie\int\frac{d\omega}{2\pi}\int \frac{d^2k_\bot}{(2\pi)^2}
\frac{k_\bot \unit \psi (i\omega \sigma - k_\bot^2)+i\sigma_\chi (\b k_\bot \omega-k_\bot^2\unit z)}{(\sigma^2+\sigma_\chi^2)(\omega-\omega_1)(\omega-\omega_2)}e^{-i\omega x_-+i \b k_\bot \cdot \b b}\label{g18}\\
&=\int\frac{d^2k_\bot}{(2\pi)^2} e^{i \b k_\bot \cdot \b b}\int_{-\infty}^{+\infty} \frac{d\omega}{2\pi} \frac{\b f(\omega)}{(\omega-\omega_1)(\omega-\omega_2)}e^{-i\omega x_-} \label{g19}\,,
\gal
where I denoted
\ball{g20}
\b f(\omega)= -\frac{ie}{\sigma^2+\sigma_\chi^2}
\left[k_\bot \unit \psi (i\omega \sigma - k_\bot^2)+i\sigma_\chi (\b k_\bot \omega-k_\bot^2\unit z)\right]\,.
\gal
 Closing the  integration contour in  \eq{g19}  by an infinite semi-circle in the lower half-plane we find
 at $x_->0$
\bal
\b B =& \int\frac{d^2k_\bot}{(2\pi)^2} e^{i \b k_\bot \cdot \b b}\frac{i}{\omega_2-\omega_1}\left[ e^{-i\omega_1 x_-}\b f(\omega_1)\theta(k_\bot-\sigma_\chi)- e^{-i\omega_2 x_-}\b f(\omega_2)\right]\theta(x_-)\,. \label{g21}
\gal
 The value of $\sigma_\chi$ probably does not exceed a few MeV at best, while typical $k_\bot$ is in the range $20-200$~MeV corresponding to $b$'s in the range $1-10$~fm. Therefore, only the case   $k_\bot^2\gg \sigma^2+\sigma_\chi^2$ has a practical significance. This allows us to approximate the poles of \eq{g17} as follows
\ball{g23}
\omega_{1,2}\approx  \frac{k_\bot^2(-i\sigma\pm \sigma_\chi)}{\sigma^2+\sigma_\chi^2}=\frac{k_\bot^2}{i\sigma\pm \sigma_\chi}\,.
\gal
Magnetic field at  $x_->0$ becomes
\ball{g25}
\b B\approx   \int\frac{d^2k_\bot}{(2\pi)^2} e^{i \b k_\bot \cdot \b b}\frac{i}{\omega_2-\omega_1}\left[ e^{-i\omega_1 x_-}\b f(\omega_1)- e^{-i\omega_2 x_-}\b f(\omega_2)\right]\,.
\gal
Its polar component  is given by 
\ball{g27}
B_\phi = \int\frac{d^2k_\bot}{(2\pi)^2} e^{i \b k_\bot \cdot \b b}\frac{i}{\omega_2-\omega_1} \hat {\b \psi}\cdot\left[ e^{-i\omega_1 x_-}\b f(\omega_1)- e^{-i\omega_2 x_-} \b f(\omega_2)\right]\,,
\gal
where $\phi$  is the angle between the impact parameter $\b b$ and the $x$-axis. Integration over the directions of $\b k_\bot$ given by the polar angle  $\psi$ is done as follows:
\ball{g28}
&\int_0^{2\pi} e^{i\b k_\bot \cdot \b b }\unit \psi \,d\psi=\int_0^{2\pi} e^{ik_\bot b \cos(\psi-\phi)}(-\unit x \sin\psi+\unit y \cos\psi) \,d\psi= 2\pi i J_1(k_\bot b)\unit \phi\,, 
\gal
Using \eq{g28} in \eq{g27} and substituting \eq{g20},\eq{g23} we have:
\ball{g29}
B_\phi&=-\int_0^\infty \frac{dk_\bot k_\bot}{2\pi}iJ_1(k_\bot b)\frac{ek_\bot}{2(\sigma^2+\sigma_\chi^2)}\left[
 (i\sigma-\sigma_\chi)e^{-i \frac{k_\bot^2x_-}{i\sigma+\sigma_\chi}}+(i\sigma +\sigma_\chi)e^{-i \frac{k_\bot^2x_-}{i\sigma-\sigma_\chi}}\right]\,.
\gal
The remaining integral can be done analytically yielding 
\ball{g31}
B_\phi=\frac{eb}{8\pi x_-^2}e^{-\frac{b^2\sigma}{4x_-}}\left[
\sigma \cos\left( \frac{b^2\sigma_\chi}{4x_-}\right)+\sigma_\chi\sin \left(\frac{b^2\sigma_\chi}{4x_-}\right)\right]\,. 
\gal
Turning to the component of magnetic field aligned along the $\b b$-direction we obtain:
\ball{g33}
B_r= \int\frac{d^2k_\bot}{(2\pi)^2} e^{i \b k_\bot \cdot \b b}\frac{i}{\omega_2-\omega_1} \unit k_\bot\cdot\left[ e^{-i\omega_1 x_-}\b f(\omega_1)- e^{-i\omega_2 x_-} \b f(\omega_2)\right]\,.
\gal
Angular integration is done using 
\ball{g32}
&\int_0^{2\pi} e^{i\b k_\bot \cdot \b  b }\unit k_\bot \,d\psi= \int_0^{2\pi} e^{ik_\bot b \cos(\psi-\phi)}(\unit x \cos\psi+\unit y \sin\psi) \,d\psi= 2\pi i J_1(k_\bot b)\unit b\,. 
\gal
Plugging the $k_\bot$-component of $\b f$ from \eq{g20} and integrating over $k_\bot$ we derive
\ball{g34}
B_r=\frac{eb}{8\pi x_-^2}e^{-\frac{b^2\sigma}{4x_-}}\left[
\sigma \sin\left( \frac{b^2\sigma_\chi}{4x_-}\right)-\sigma_\chi\cos \left(\frac{b^2\sigma_\chi}{4x_-}\right)\right]\,. 
\gal
Finally, repeating the by now familiar procedure and using the integral 
\ball{g36}
&\int_0^{2\pi}e^{i\b k_\bot \cdot \b b }\unit z\,d\psi = 2\pi J_0(k_\bot b)\unit z
\gal
we find for the longitudinal field component:
\ball{g38}
B_z= -\frac{e}{4\pi x_-}e^{-\frac{b^2\sigma}{4x_-}}\left[
\sigma \sin\left( \frac{b^2\sigma_\chi}{4x_-}\right)-\sigma_\chi\cos \left(\frac{b^2\sigma_\chi}{4x_-}\right)\right]\,.
\gal
It is seen in \eq{g34} and \eq{g38} that the  field components $B_r$ and $B_z$ are generated only at a finite chiral conductivity $\sigma_\chi$. 

\begin{figure}[ht]
\begin{tabular}{cc}
      \includegraphics[height=5cm]{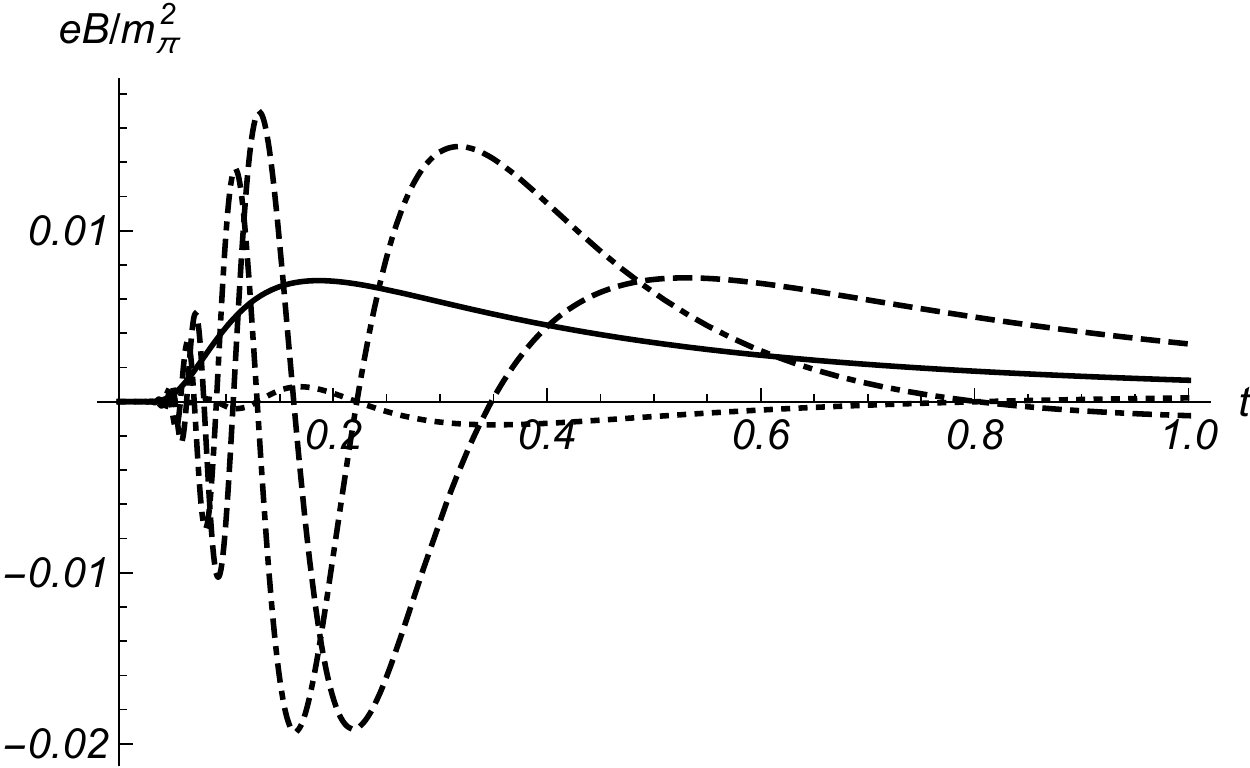} &
      \includegraphics[height=5cm]{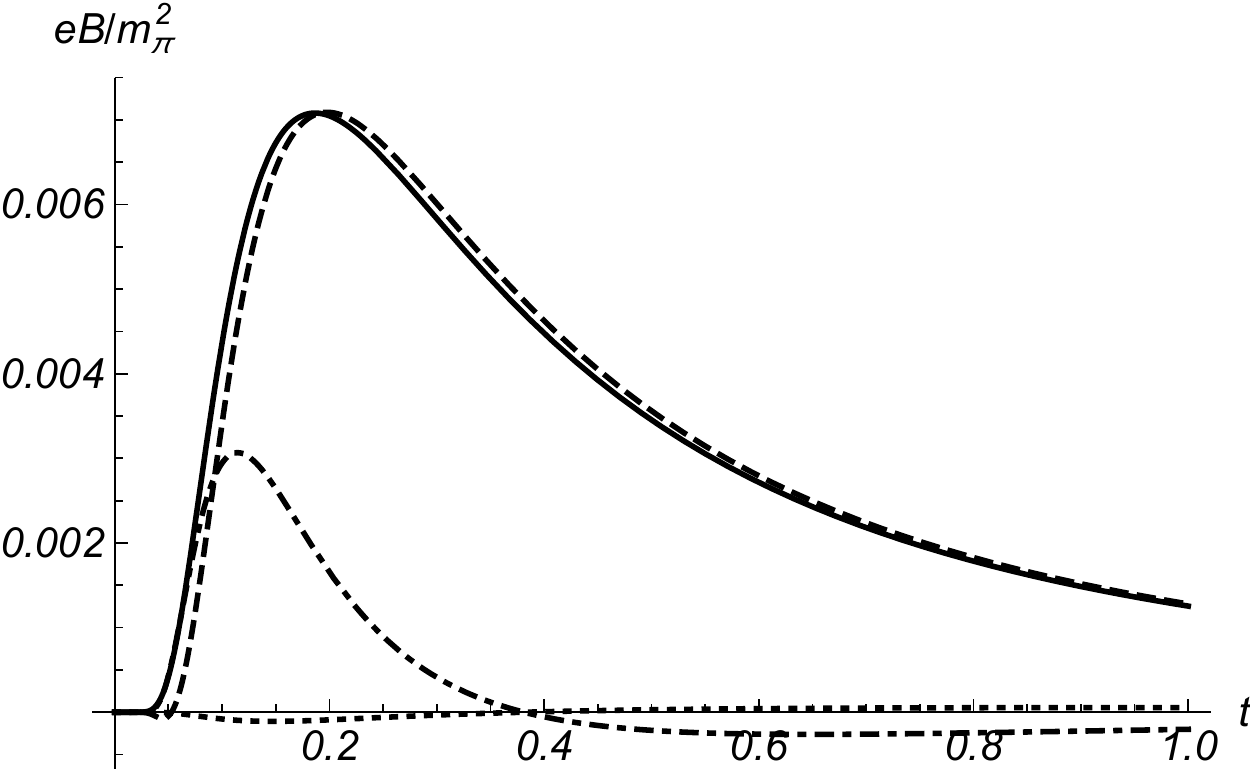}
            \end{tabular}
  \caption{Magnetic field of a point charge as a function of time $t$ at $z=0$. (Free space contribution is not shown). Electrical conductivity $\sigma= 5.8$~MeV. Solid line on both panels corresponds to $B=B_\phi$ at $\sigma_\chi=0$. Broken lines correspond to $B_\phi$ (dashed), $B_r$ (dashed-dotted) and $B_z$ (dotted) with $\sigma_\chi = 15$~MeV on the left panel and $\sigma_\chi=1.5$ MeV on the right panel. Note that the vertical scale on the two panels is different. 
  }
\label{fig2}
\end{figure}

Eqs.~\eq{g31},\eq{g33} and \eq{g38} is the main result of this paper. It shows that at finite $\sigma_\chi$, magnetic field of a point charge acquires two components that are absent in the chirally neutral medium: the radial and the longitudinal components. All field components oscillate at early times. This is clearly seen in \fig{fig2}.  The $B_z$ and $B_r$ components change sign at light-cone times 
\ball{g43} 
x_{-}^{(n)}=\frac{b^2\sigma_\chi}{4[\arctan\frac{\sigma_\chi}{\sigma}+\pi n]}\,, \quad n=0,1,\ldots\,,
\gal
while the $B_\phi$ components changes sign at 
\ball{g44} 
 \tilde x_-^{(n)}=\frac{b^2\sigma_\chi}{4[-\arctan\frac{\sigma}{\sigma_\chi}+\pi n]}\,, \quad n=0,1,\ldots\,,
\gal
The latest oscillation corresponds to $n=0$; it increases with $\sigma_\chi$.

\section{Discussion and summary}\label{sec:i}

 We discussed the chiral topological effect on electromagnetic field in the Quark-Gluon Plasma. In our model the anomalous current density  is given by $\b j= \sigma_\chi\b B$ with constant chiral conductivity $\sigma_\chi$. For the energy and time scales of the QGP this model gives a reasonable physical picture of the electromagnetic field space-time evolution.   There are two major results presented in this paper.

(i) I showed that solutions to the Maxwell equations are not stable  in the presence of the chirality imbalance. It is possible that electromagnetic field collapses into a set of magnetic knots. This problem certainly deserves a dedicated study and may be important in cosmology.   However, as far as  heavy-ion collisions are concerned, this instability has negligible  impact on the QGP because it originates from soft modes $k<\sigma_\chi$ that do not exist in the QGP of realistic dimensions. The maximal growth rate of unstable modes is $\left(\sqrt{\sigma^2+\sigma_\chi^2}-\sigma\right)/2$.

(ii) I derived an analytical expression for magnetic field produced by valence charges in quark-gluon plasma at finite chiral conductivity $\sigma_\chi$. Its components are given by equations \eq{g31},\eq{g34} and \eq{g38}, which  indicate emergence of the radial $B_r$ and  longitudinal $B_z$ components of magnetic field (as compared to the $\sigma_\chi=0$ case). If $\sigma_\chi$ is not much smaller than $\sigma$, then all components perform oscillations at early times after the collision. Since magnetic field is strongest at early times, these oscillations should have important impact on heavy-ion phenomenology. In particular, they may weaken effects that depend on the magnetic field direction, such as the $B$-dependent elliptic flow \cite{Tuchin:2011jw,Mohapatra:2011ku} and charge separation effect \cite{Kharzeev:2007jp}. This is especially true for the charge separation effect that requires sufficiently large $\sigma_\chi$.

 In this paper, I considered the simplest model that incorporates the chiral anomaly in electrodynamics. Its main advantages are  that it describes the experimentally observable charge separation in heavy-ion collisions and can be solved  analytically. However, it has serious drawbacks as well:  chiral conductivity of a realistic plasma is a complicated function of space and time.  Thus, the main outstanding problem is to find a more realistic model for the chiral anomaly and verify which of the above results, and to what extent, survive in an improved formulation. This can serve as a benchmark for the full magneto-hydrodynamical treatment of the problem.

\acknowledgments
I  am grateful to Dmitri Kharzeev for an informative discussion an comments on a draft version of this manuscript. I would like to thank Qun Wang for pointing out a mistake in Eq.~(62) in an earlier version of this paper. 
This work  was supported in part by the U.S. Department of Energy under Grant No.\ DE-FG02-87ER40371.


\end{document}